\documentclass[10pt,twocolumn,reprint,amsmath,amssymb,aps,pra,showpacs,longbibliography,superscriptaddress]{revtex4-1}
\usepackage[latin9]{inputenc}
\usepackage{amsmath}
\usepackage{amssymb}
\usepackage{graphicx}

\makeatletter
\usepackage{amsfonts}
\usepackage{graphicx}
\usepackage{graphics}

\begin{document}
\title{Microscopic many-body theory of two-dimensional coherent spectroscopy of exciton-polarons in one-dimensional materials}
\author{Jia Wang, Hui Hu, Xia-Ji Liu}
\address{Centre for Quantum Technology Theory, Swinburne University of Technology,
Melbourne 3122, Australia}

\begin{abstract}
We have developed a microscopic many-body theory of two-dimensional coherent spectroscopy (2DCS) for a model of polarons in one-dimensional (1D) materials. Our theory accounts for contributions from all three processes: excited-state emission (ESE), ground-state bleaching (GSB), and excited-state absorption (ESA). While the ESE and GSB contributions can be accurately described using a Chevy's ansatz with one particle-hole excitation, the ESA process requires information about the many-body eigenstates involving two impurities. To calculate these double polaron states, we have extended the Chevy's ansatz with one particle-hole excitation. The validity of this ansatz was verified by comparing our results with an exact calculation using Bethe's ansatz. Our numerical results reveal that in the weak interaction limit, the ESA contribution cancels out the total ESE and GSB contributions, resulting in less significant spectral features. However, for strong interactions, the features of the ESA contribution and the combined ESE and GSB contributions remain observable in the 2DCS spectra. These features provide valuable information about the interactions between polarons. Additionally, we have investigated the mixing time dynamics, which characterize the quantum coherences of the polaron resonances. Overall, our theory provides a comprehensive framework for understanding and interpreting the 2DCS spectra of polarons in 1D materials, shedding light on their interactions and coherent dynamics.
\end{abstract}
\maketitle

\section{Introduction}

Monolayer transition-metal-dichalcogenides (TMD) have attracted intense interest due to their intriguing electrical and optical properties in low-dimensions arising from tightly Coulomb-bound electron-hole
pairs (i.e., excitons) and excitonic complexes such as trions and bi-excitons \cite{Wang2018RMP,Berkelbach2018}. From a perspective of developing practical applications of low-energy-threshold electronics and optoelectronics, it is of central importance to characterize and manipulate the nonlinearity or many-particle interactions among excitons and excitonic complexes \cite{Hu2020PRA}. In this respect, a remarkable tool is the nonlinear two-dimensional coherent spectroscopy (2DCS) built on the four-wave-mixing \cite{Jonas2003,Li2006,Cho2008,Hao2016}. It measures the full third-order nonlinear optical susceptibility of materials as functions of excitation and emission energies, and can be implemented to probe the formation and dynamics of excitons and excitonic complexes at the femtosecond timescale. In recent measurements, 2DCS has been successfully applied to characterize a novel excitonic complex known as exciton-polarons in monolayer MoSe$_{2}$ \cite{Huang2023}, which is formed by doping the two-dimensional materials with electrons (or holes) \cite{Sidler2017,Efimkin2017}. For the first time, it has also been used to reveal the interaction effect between exciton-polarons in monolayer WSe$_{2}$ \cite{Muir2022}.

Here, we aim to present a microscopic many-body theory for 2DCS of exciton-polarons in TMD materials when their motion is restricted to a specific direction. Our motivation of investigating such effectively one-dimensional (1D) materials is two-fold. First, the deterministic dimensionality engineering of TMD materials brings additional benefit of coherent propagation of light-emitting quasipartcles for constructing all-optical integrated logic circuits. As experimentally demonstrated most recently by Chernikov and Menon and their co-workers \cite{Dirnberger2021}, this can be achieved by merging 1D semiconductor nanowires with TMD monolayers into hybirid heterostructure, where the strain mismatch confines the motion of excitons and creates an artificial strain-induced exciton transport channel. By narrowing the channel width down to about 60 nanometers in the near future \cite{Dirnberger2021}, the truly 1D regime could be reached to demonstrate mesoscopic quantum transport of excitons and excitonic complexes.

On the other hand, a microscopic theoretical framework of 2DCS of 1D quantum many-body systems might become feasible, because of the much-reduced numerical workload for describing the excited many-body states involved in 2DCS. In the past, theoretical descriptions of 2DCS rely heavily on the simplification that treats the many-body interacting system as a few-energy-level structure \cite{Li2006,Cho2008,Hao2016}. Only recently, microscopic many-body descriptions have been developed to study the time-resolved ARPES spectrum of a two-band model semiconductor \cite{Stefanucci2021PRB} and attempted to understand the 2DCS of exciton-polarons in TMD materials \cite{Tempelaar2019,Lindoy2022,Wang2023,Wang2022arXiv,Hu2022arXiv,Hu2023AB,Wang2022Review}. For the standard rephasing mode \cite{Cho2008,Hao2016}, the two contributions from the excited-state emission (ESE) process and the ground-state bleaching (GSB) process have been numerically calculated \cite{Hu2022arXiv,Hu2023AB}. The remaining process of the excited-state absorption (ESA) is often neglected, partly because the related two-polaron states are difficult to account for due to their enormously large Hilbert space.

In this work, we overcome such a difficulty by constructing approximate but reasonably accurate two-polaron states in one dimension. The account of all the three processes completes the full microscopic description of the rephasing 2DCS. Therefore, our results pave a useful way to quantitatively understand the 2DCS of exciton-polarons in strain-engineered 1D TMD materials, to be carried out in the near future. By extending our calculations to the two-dimensional case with a restricted Hilbert space, the interaction effect and dynamics of exciton-polarons, as recently observed in monolayer WSe$_{2}$ \cite{Muir2022}, may also be better explained.

The rest of the paper is organized as follows. In the following section, we outline the overall theoretical approach. We discuss the model Hamiltonian and derive the detailed expressions for the three processes of the rephasing 2DCS, with the help of Chevy's ansatz that takes into account one-particle-hole excitations for the perturbed Fermi sea \cite{Chevy2006}. In section 3, we first present the numerical results for both the single polaron state and the bipolaron state. The accuracy of Chevy ansatz is examined, in comparison with the available exact solution based on the Bethe ansatz \cite{Guan2013,Huber2019} for the specific interaction parameter. We then discuss in detail the 2DCS at the zero mixing time delay $t_{2}=0$ and show the prominent feature due to interaction effect between two polarons. The coherent dynamics as a function of nonzero mixing time delay $t_{2}$ is also investigated. In section 4, we briefly summarize the key results of the work. Finally, Appendix A gives the matrix elements used to diagonalize the model Hamiltonian and Appendix B presents the finite size scaling for the bipolaron energy and the energy difference responsible for the induced interaction between polarons.

\section{Theoretical Approach}

\subsection{Model Hamiltonian}
We consider a system of excitons interacting with surrounding excess charges of electrons or holes in a 1D space. The excess charges are represented by the fermionic creation and annihilation field operators $c_k^\dagger$ and $c_k$, respectively. The excitons, characterized by a significantly larger binding energy compared to other energy scales in the system, are described by the bosonic creation and annihilation field operators $X_k^\dagger$ and $X_k$, respectively, with their internal degrees of freedom frozen \cite{Wang2018}. The system can be effectively described by the following Hamiltonian \cite{Efimkin2017}:
\begin{eqnarray}
\mathcal{H} &=& \sum_{k}\left[\epsilon_{k} c_{k}^{\dagger} c_{k}+\epsilon_{k}^X X_{k}^{\dagger} X_{k}\right]+U \sum_{qkp} X_{k}^{\dagger} c_{q-k}^{\dagger} c_{q-p} X_{p} \nonumber \\
&&+ U_X \sum_{qkp} X_{k}^{\dagger} X_{q-k}^{\dagger} X_{q-p} X_{p}, \label{eq:Hamiltonian}
\end{eqnarray}
where $U$ represents the interaction strength between excitons and excess charges, and $U_X$ represents the bare interaction strength between excitons. $\epsilon_{k}^X$ and $\epsilon_{k}$ represent the kinetic energies of excitons and excess charges, respectively. It is convenient to set the reduced Planck constant $\hbar = 1$.  

One can verify that the Hamiltonian $\mathcal{H}$, the total momentum $P=\sum_k k \left(c_{k}^{\dagger} c_{k}+ X_{k}^{\dagger} X_{k}\right)$, the number of excitons $N_X=\sum_k X_{k}^{\dagger} X_{k}$, and the excess charge number $N=\sum_k c_k^\dagger c_k$ all commute with each other. This implies that the Hamiltonian is block-diagonal with respect to the quantum numbers $P$, $N$, and $N_X$. In the subsequent analysis, we will focus on the case where $P=0$ and a fixed number of excess charges $N$. We denote the block of the Hamiltonian with exciton number $N_X$ as $\mathcal{H}_{N_X}$. To obtain the matrix representation of $\mathcal{H}_{N_X}$, we can expand $\mathcal{H}$ using a basis set that includes $N_X$ excitons. In this study, we employ an extended version of the Chevy's ansatz \cite{Chevy2006} to construct such a basis set, which will be described in detail later. This basis set allows us to represent the many-body states with a specific number of excitons, enabling us to diagonalize $\mathcal{H}_{N_X}$ and study the properties of the system. 

Furthermore, we consider a zero-temperature scenario, where the initial state is prepared as the ground state $\left|\textrm{FS}\right\rangle = \sum_{\epsilon_k \le E_F} c_k^\dagger c_k | {\rm vac} \rangle$, corresponding to the Fermi sea with no excitons. The excess charges are described by occupying all single-particle states below the Fermi energy $E_F$. Here, $|\textrm{vac}\rangle$ represents the vacuum state. The number of excess charges $N$ is fixed by the Fermi energy $E_F$. It is common and convenient to define the zero-point energy by subtracting the background energy of the Fermi sea, denoted as $E_{\rm FS} = \sum_{\epsilon_k\le E_F} \epsilon_k$, from the total Hamiltonian $\mathcal{H}$. By setting $E_{\rm FS}$ as the zero-point energy reference, we can redefine the Hamiltonian as $\mathcal{H} \rightarrow \mathcal{H} - E_{\rm FS}$ without affecting the physical properties and dynamics of the system. 

\subsection{Two-dimensional coherent spectroscopy}
\begin{figure}
\begin{centering}
\includegraphics[width=0.98\columnwidth]{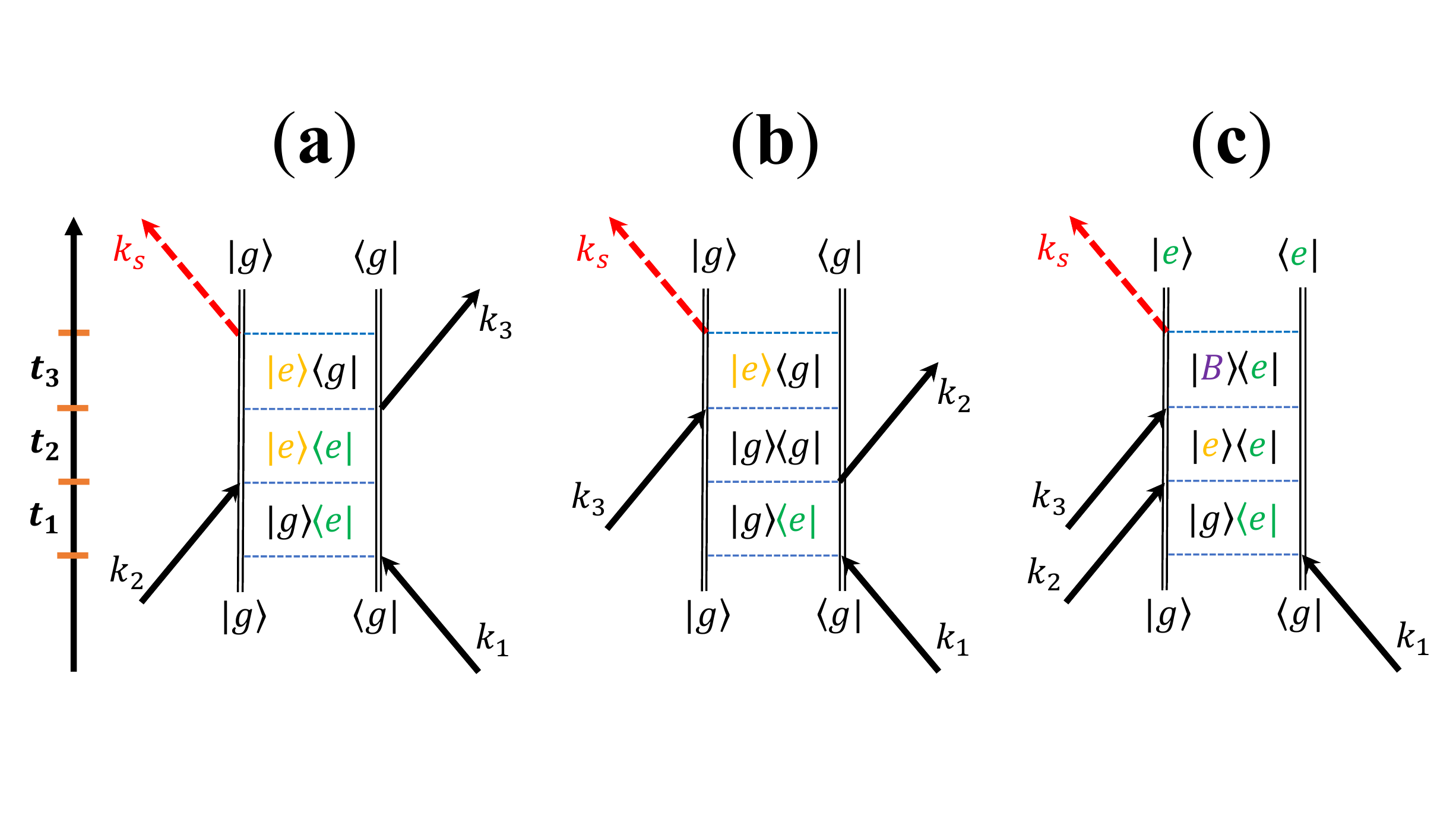}
\caption{\label{fig:2DCS_sketch} Three double-sided Feynman diagrams that represent the three contributions to the standard rephasing 2D coherent spectra under the phase-match condition $\mathbf{k}_{s}=-\mathbf{k}_{1}+\mathbf{k}_{2}+\mathbf{k}_{3}$, with the time ordering of excitation pulses indicated on the left \cite{Cho2008,Hao2016}. The evolution, mixing, and emission time delays are labeled as $t_{1}$, $t_{2}$, and $t_{3}$, respectively. (a) shows the process of excited-state emission (ESE), $R_{2}(t_{1},t_{2},t_{3})$, (b) corresponds to the ground-state bleaching (GSB), $R_{3}(t_{1},t_{2},t_{3})$, and (c) gives the excited-state absorption (ESA), $R_{1}^{*}(t_{1,}t_{2},t_{3})$. In the diagrams, we use $\left|g\right\rangle $ to denote the Fermi sea and $\left|e\right\rangle $ ($\left|B\right\rangle $) to label the many-body states with a single exciton (two excitons or a bi-exciton), respectively. There are infinitely many many-body states $\left|e\right\rangle $ (Fermi polaron) and $\left|B\right\rangle $ (bi-polarons), as indicated by different colors.}
\end{centering}
\end{figure}

2DCS spectroscopy has been implemented in experiments to study exciton-polaron physics in TMD materials \cite{Hao2016,Huang2023,Muir2022}. In 2DCS, three excitation pulses with momentum $\mathbf{k}_{1}$, $\mathbf{k}_{2}$, and $\mathbf{k}_{3}$ are applied to the system being studied at times $\tau_1$, $\tau_{2}$, and $\tau_{3}$, separated by an evolution time delay $t_{1}=\tau_{2}-\tau_{1}$ and a mixing time delay $t_{2}=\tau_{3}-\tau_{2}$, as shown in the left part of Figure \ref{fig:2DCS_sketch}. These pulses generate a signal with momentum $\mathbf{k}_{s}$ as a result of the nonlinear third-order process of the many-body interaction effect. The signal can then be measured after an emission time delay $t_{3}$ using frequency-domain heterodyne detection.

During the excitation period, each excitation pulse creates or annihilates an exciton. Since the photon momentum of the excitation pulses is negligible, the exciton has zero momentum. Therefore, each pulse can be described by the interaction operator $V$, given by
\begin{equation}
V\propto X_{0}+X_{0}^{\dagger}.
\end{equation}
The excitation pulse periods are much shorter than the time delays and are therefore assumed to be instantaneous. The form of $V$ ensures that the eigenstates of $\mathcal{H}_{N_X}$ only couple to the eigenstates of $\mathcal{H}_{N_X\pm1}$ after each pulse. 
According to the standard nonlinear response theory \cite{Cho2008}, the signal is given by the third-order nonlinear response function:
\begin{equation}
\mathcal{R}^{(3)}\propto\left\langle \left[\left[\left[V\left(t_{1}+t_{2}+t_{3}\right),V\left(t_{1}+t_{2}\right)\right],V\left(t_{1}\right)\right],V\right]\right\rangle,
\end{equation}
where $V(t)\equiv e^{i\mathcal{H}t}Ve^{-i\mathcal{H}t}$ represents the time-dependent interaction operator, and $\left\langle \cdots\right\rangle $ denotes the quantum average over the initial many-body configuration of the system without excitation pulses, which, at zero temperature, corresponds to the ground state. By expanding the three bosonic commutators, four distinct correlation functions and their complex conjugates are obtained \cite{Cho2008}. In the rephasing mode, with $t_{1}>0$ and $\mathbf{k}_{s}=-\mathbf{k}_{1}+\mathbf{k}_{2}+\mathbf{k}_{3}$, three relevant contributions are dominant. These contributions include (A) Excited-State Emission (ESE) process:
\begin{equation}
R_{2}=\left\langle VV\left(t_{1}+t_{2}\right)V\left(t_{1}+t_{2}+t_{3}\right)V\left(t_{1}\right)\right\rangle,
\end{equation}
visualized by the double-sided Feynman diagram in Fig. \ref{fig:2DCS_sketch}(a); (B) Ground-State Bleaching (GSB) process:
\begin{equation}
R_{3}=\left\langle VV\left(t_{1}\right)V\left(t_{1}+t_{2}+t_{3}\right)V\left(t_{1}+t_{2}\right)\right\rangle,
\end{equation}
visualized by the double-sided Feynman diagram in Fig. \ref{fig:2DCS_sketch}(b); and (C) Excited-State Absorption (ESA) process:
\begin{equation}
R_{1}^{*}=-\left\langle VV\left(t_{1}+t_{2}+t_{3}\right)V\left(t_{1}+t_{2}\right)V\left(t_{1}\right)\right\rangle, \label{eq:R1sDEF}
\end{equation}
involving intermediate many-body states of two excitons, as shown in Fig. \ref{fig:2DCS_sketch}(c).

A recent microscopic calculation of the ESE and GSB contributions has been conducted \cite{Hu2022arXiv,Hu2023AB}. Following the approach outlined in Ref. \cite{Hu2022arXiv,Hu2023AB}, we can derive the expressions:
\begin{eqnarray}
R_{2} & = & \sum_{nm}Z_{n}Z_{m}e^{i\mathcal{E}_n^{(P)} t_{1}}e^{i\left[\mathcal{E}_n^{(P)}-\mathcal{E}_m^{(P)}\right]t_{2}}e^{-i\mathcal{E}_m^{(P)}t_{3}},\\
R_{3} & = & \sum_{nm}Z_{n}Z_{m}e^{i\mathcal{E}_n^{(P)} t_{1}}e^{-i\mathcal{E}_m^{(P)}t_{3}},
\end{eqnarray}
where the indices $n$ and $m$ span the entire set of many-body eigenstates of the Hamiltonian $\mathcal{H}_1$, which corresponds to the case where the exciton number is fixed to $N_X=1$. The Hamiltonian $\mathcal{H}_1$ can be interpreted as a polaron Hamiltonian, and its eigenenergies and eigenstates are denoted as $\mathcal{E}_n^{(P)}$ and $\left| n \right\rangle_P$, respectively. The residue $Z_n$ is given by $Z_n=|\phi^{(n)}_0|^2$, where $\phi^{(n)}_0 \equiv \langle {\rm FS} | X_0 |n\rangle_P$ represents the projection of the polaron state onto the Fermi sea with a non-interacting impurity at zero momentum. It satisfies $\sum_n |\phi_0^{(n)}|^2 = \langle {\rm FS}| X_0 X^\dagger_0 | {\rm FS} \rangle =1$.

To calculate the ESA contribution, we insert $V\propto X_{0}+X_{0}^{\dagger}$ into Eq. (\ref{eq:R1sDEF}) and obtain,
\begin{equation}
R_{1}^{*}=-\left\langle \textrm{FS}\right|X_{0}e^{i\mathcal{H}_{1}\left(t_{1}+t_{2}+t_{3}\right)}X_{0}e^{-i\mathcal{H}_{2}t_{3}}X_{0}^{\dagger}e^{-i\mathcal{H}_{1}t_{2}}X_{0}^{\dagger}\left|\mathrm{FS}\right\rangle,\label{eq:R1s}
\end{equation}
where $\mathcal{H}_{1}$ and $\mathcal{H}_{2}$ are the Hamiltonians for a single exciton and two excitons (bi-polaron), respectively. The many-body eigenenergies and eigenstates of $\mathcal{H}_{2}$ are denoted as ${\mathcal E}_\eta^{(B)}$ and $\left| \eta \right\rangle_B$, respectively. For simplicity, we will use Roman letters to represent the polaron eigenstates and eigenenergies, and Greek letters for the bi-polaron eigenstates and eigenenergies. Hereafter, the superscripts and subscripts $P$ and $B$ are dropped for convenience. By applying $e^{i \mathcal{H}_1 t} = \sum_n |n\rangle \langle n | e^{i \mathcal{E}_n t}$ and $e^{i \mathcal{H}_2 t} = \sum_\eta |\eta\rangle \langle \eta | e^{i \mathcal{E_\eta} t}$, we can simplify the expression for $R_1^{*}$ as follows:
\begin{equation}
R_{1}^{*}=-\sum_{nm\eta}\phi_{0}^{(n)}\Phi^{(n,\eta)}\Phi^{(m,\eta)*}\phi_{0}^{(m)*}e^{i\mathcal{E}_n t_{1}}e^{i\left[\mathcal{E}_n-\mathcal{E}_m\right]t_{2}}e^{i\left[\mathcal{E}_n-\mathcal{E}_{\eta}\right]t_{3}},
\end{equation}
where $\Phi^{(n,\eta)} \equiv \langle n| X_0 | \eta \rangle$ represents the projection of the $\eta$-th bi-polaron state onto the $n$-th polaron state with a non-interacting impurity at zero momentum. This expression can be simply understood from the double-sided Feynman diagrams shown in Fig. \ref{fig:2DCS_sketch}(c). The four weighting factors represent the transition rates and the three exponents show the dynamical phases accumulated during the three time-evolution periods.

We can also define $\Phi_0^{(\eta)} \equiv  \langle \textrm{FS} | X_0X_0 | \eta \rangle/\sqrt{2}$ as the projection of the bi-polaron state onto the Fermi sea with two non-interacting impurities, with the factor $1/\sqrt{2}$ arising from the normalization condition. It can be verified that by inserting $\sum_n |n\rangle \langle n| = 1$ between the two $X_0$ operators, we have $\Phi_0^{(\eta)} =  \sum_n \phi_0^{(n)}\Phi^{(n,\eta)}/\sqrt 2$. Furthermore, we have $\sum_{nm\eta}\phi_{0}^{(n)}\Phi^{(n,\eta)}\Phi^{(m,\eta)*}\phi_{0}^{(m)*} =2 \sum_\eta |\Phi_0^{(\eta)}|^2 = 2$. 

After performing a double Fourier transformation with respect to $t_{1}$ and $t_{3}$ on $R_{2}(t_{1},t_{2},t_{3})$, $R_{3}(t_{1},t_{2},t_{3})$, and $R_{1}^{*}(t_{1},t_{2},t_{3})$, we obtain the 2DCS spectrum given by the following expressions:
\begin{eqnarray}
\mathcal{S}_{\mathrm{ESE}}\left(\omega_{1},t_{2},\omega_{3}\right) & = & -\sum_{nm}\frac{Z_{n}Z_{m}}{\omega_{1}^{+}+\mathcal{E}_n}\frac{e^{i\left[\mathcal{E}_n-\mathcal{E}_m\right]t_{2}}}{\omega_{3}^{+}-\mathcal{E}_m},\label{eq:S_ESE}\\
\mathcal{S}_{\mathrm{GSB}}\left(\omega_{1},t_{2},\omega_{3}\right) & = & -\sum_{nm}\frac{Z_{n}Z_{m}}{\omega_{1}^{+}+\mathcal{E}_n}\frac{1}{\omega_{3}^{+}-\mathcal{E}_m},\label{eq:S_GSB}\\
\mathcal{S}_{\mathrm{ESA}}\left(\omega_{1},t_{2},\omega_{3}\right) & = & \sum_{nm\eta}\frac{\phi_{0}^{(n)}\Phi^{(n,\eta)}}{\omega_{1}^{+}+\mathcal{E}_n}e^{i\left[\mathcal{E}_n-\mathcal{E}_m\right]t_{2}}\frac{\Phi^{(m,\eta) *}\phi_{0}^{(m)*}}{\omega_{3}^{+}-\mathcal{E}_{\eta}+\mathcal{E}_n}.\label{eq:S_ESA}
\end{eqnarray}
Here, $\omega^+ \equiv \omega + i0^+$ represents the complex frequency with a positive infinitesimal imaginary part. The spectra $\mathcal{S}_{\mathrm{ESE}}$, $\mathcal{S}_{\mathrm{GSB}}$, and $\mathcal{S}_{\mathrm{ESA}}$ depend on the excitation energy $\omega_1$, the mixing time $t_2$, and the emission energy $\omega_3$. 

In the limit of vanishing interaction strength between excitons and the Fermi sea ($U = U_X = 0$), only the lowest energy levels in the polaron states and bipolaron states contribute significantly to the 2DCS. Specifically, we have $n = \eta = 0$, $|n= 0\rangle = X_0^\dagger |{\rm FS}\rangle$, and $|\eta= 0\rangle = X_0^\dagger X_0^\dagger|{\rm FS}\rangle/\sqrt{2}$. In this case, we find that $\mathcal{E}_n = \mathcal{E}_\eta = 0$, $\phi_{0}^{(n=0)} = 1$, and $\Phi^{(n=0,\eta=0)} = \sqrt{2}$. As a result, the expressions for the 2DCS spectra simplify to $\mathcal{S}_{\rm{ESE}}(\omega_1, t_2, \omega_3) = \mathcal{S}_{\rm{GSB}}(\omega_1, t_2, \omega_3) = -1/(\omega_1^+\omega_3^+)$ and $\mathcal{S}_{\rm{ESA}}(\omega_1, t_2, \omega_3) = 2/(\omega_1^+\omega_3^+)$. Therefore, the total 2DCS spectrum, obtained by summing these contributions, adds up to zero, as anticipated for a non-interacting system. This result reflects the absence of interactions and signifies that the non-interacting system does not exhibit any coherent two-dimensional spectroscopic features.

Furthermore, we can observe the following integrals over the frequency variables: $\int\int d\omega_1 d\omega_3 \mathcal{S}_{\mathrm{ESE}} = \sum_{nm}Z_nZ_m e^{i[\mathcal E_n-\mathcal E_m]t_2}$, $\int\int d\omega_1 d\omega_3 \mathcal{S}_{\mathrm{GSB}} = \sum_{nm}Z_nZ_m = 1$, and $\int\int d\omega_1 d\omega_3 \mathcal{S}_{\mathrm{ESA}}  = -\sum_{nm\eta}\phi_{0}^{(n)}\Phi^{(n,\eta)}\Phi^{(m,\eta)*}\phi_{0}^{(m)*} e^{i[\mathcal E_n - \mathcal E_m]t_2}$. Applying $\sum_\eta \Phi^{(n,\eta)}\Phi^{(m,\eta)*}= \langle n | X_0 X_0^\dagger | m \rangle = \delta_{nm} + \langle n | X_0^\dagger X_0 | m \rangle = \delta_{nm} + \phi_0^{(n)*}\phi_0^{(m)}$ gives 
\begin{equation}
\int\int d\omega_1 d\omega_3 \mathcal{S}_{\mathrm{ESA}}  = - \int\int d\omega_1 d\omega_3 \mathcal{S}_{\mathrm{ESE}} - \int\int d\omega_1 d\omega_3 \mathcal{S}_{\mathrm{GSB}}. 
\end{equation}
This result indicates that the total signal integrated over all frequencies is exactly zero. These general conclusions hold universally and are not contingent on the microscopic details or the variational basis employed.

\subsection{Chevy's ansatz}
To solve the model Hamiltonian with general interaction strengths $U$ and $U_X$, we utilize Chevy's ansatz \cite{Chevy2006}, which allows for up to one particle-hole excitation in the Fermi sea. For the case of a single exciton, the ansatz is given by:
\begin{equation}
|n\rangle=\phi_0^{(n)} \left|0\right\rangle_1+\sum_{k_pk_h} \phi_{k_pk_h}^{(n)} \left| k_p k_h\right\rangle_1, \label{eq:PolaronAnsatz}
\end{equation}
where the basis states are defined as $\left|0\right\rangle_1 = X_0^{\dagger}|\mathrm{FS}\rangle$ and $\left| k_p k_h\right\rangle_1 = X_{-k_p+k_h}^{\dagger}c_{k_p}^\dagger c_{k_h}\left|{\rm FS}\right\rangle$. The subscript $1$ indicates the presence of a single impurity. By computing the matrix elements of ${\mathcal H}_1$ as $_1\left\langle i \right| {\mathcal H} \left| j \right \rangle_1$, where the indices $i$ and $j$ can be either $\{0\}$ or $\{k_p,k_h\}$, we can determine the variational coefficients $\phi_0^{(n)}$, $\phi_{k_pk_h}^{(n)}$, and the corresponding eigenenergy ${\mathcal E}_n$ by diagonalizing ${\mathcal H}_1$.

For the case of two excitons, the ansatz is given by:
\begin{equation}
\left|\eta\right\rangle =\sum_{k}'\Phi_{k}^{(\eta)} \left| k \right \rangle_2 +\sum_{k}'\sum_{k_{p}k_{h}}\Phi_{kk_{p}k_{h}}^{(\eta)} \left| k k_p k_h\right\rangle_2, \label{eq:BipolaronAnsatz}
\end{equation}
where the prime in the summation over $k$ indicates the avoidance of double counting (with details presented in Appendix A). The basis states are defined as $\left| k \right \rangle_2 = C_k X_{-k}^{\dagger}X_{k}^{\dagger}\left|\textrm{FS}\right\rangle$ and $\left| k k_p k_h\right\rangle_2 = C_{k,k_p,k_h}X_{-k}^{\dagger}X_{k-k_{p}+k_{h}}^{\dagger}c_{k_{p}}^{\dagger}c_{k_{h}}\left|\textrm{FS}\right\rangle$, where $C_k=1/\sqrt{1+\delta_{k,0}}$ and $C_{k,k_p,k_h}=1/\sqrt{1+\delta_{-k,k-k_p+k_h}}$ with $\delta_{ab}$ being Kronecker delta function are introduced to ensure the orthogonality of the basis. Similarly to the single impurity case, we construct the Hamiltonian matrix ${\mathcal H}_2$ with matrix elements $_2\left\langle i \right| {\mathcal H} \left| j \right \rangle_2$, where the indices $i$ and $j$ correspond to $\{k\}$ or $\{k,k_p,k_h\}$, and we find the variational coefficients $\Phi_{k}^{(\eta)}$, $\Phi_{kk_{p}k_{h}}^{(\eta)}$, and the corresponding ${\mathcal E}_\eta$ by diagonalizing ${\mathcal H}_2$. The details of the matrix elements of $\mathcal H_1$ and $\mathcal H_2$ can be found in Appendix A, where explicit expressions for these elements are provided.

The variational coefficients and eigenvalues are then utilized to obtain the expressions of Eqs. (\ref{eq:S_ESE}), (\ref{eq:S_GSB}) and (\ref{eq:S_ESA}). Specifically, $\Phi^{(n,\eta)}$ in Eq. (\ref{eq:S_ESA}) is given by $\Phi^{(n,\eta)}=\sqrt{2}\phi_0^{(n)*}\Phi_0^{(\eta)}+\sum_{k_p,k_h>k_p}\phi_{k_pk_h}^{(n)*}\Phi_{0k_pk_h}^{(\eta)}+\sum_{k_p,k_h<k_p}\phi_{k_pk_h}^{(n)*}\Phi_{-k_p+k_hk_pk_h}^{(\eta)}$. 

\section{Results and discussions}

\subsection{Numerical method}
To perform numerical investigations, we consider the Hamiltonian in Eq. (\ref{eq:Hamiltonian}) within a 1D tight-binding model consisting of $L$ sites. The excess charge density is given by $n=N/La$, where $a$ is the lattice spacing. Both the excess charges and the exciton move on the same lattice with hopping strengths $t_c$ and $t_d$, respectively. The single-particle energy dispersion relations are given by:
\begin{eqnarray}
\epsilon_{k} &=&-2 t_c \cos \left(ka\right) \simeq-2 t_c+\frac{k^2}{2 m_c}, \\
\epsilon_{k}^X &=&-2 t_d \cos \left(ka\right) \simeq-2 t_d+\frac{k^2}{2 m_X},
\end{eqnarray}
where $m_c \equiv 1/(2 t_c a^2)$ and $m_X \equiv 1/(2t_d a^2)$ in the dilute limit $(n\rightarrow 0)$ of interest. From now on, the lattice spacing is set to unity ($a=1$) unless otherwise specified. For convenience, we introduce the dimensionless quantities $u = U m_c/n$ and $u_X = U_X m_c/n$. Additionally, we define an energy unit $\epsilon_c = n^2/2m_c$. We typically assume periodic boundary conditions, which restrict the momentum $k$ on the lattice to take values within the first Brillouin zone, i.e., $k=2\pi \nu /L$ with the integer $\nu = -L/2+1,...,-1,0,1,...,L/2$. Any momentum index appearing in Eqs. (\ref{eq:PolaronAnsatz}) and (\ref{eq:BipolaronAnsatz}), therefore, should also be projected into the first Brillouin zone.

In this finite-size lattice model, the level spacing in the single-particle dispersion relation is on the order of $t_c/L$, which tends to zero as the system size $L$ approaches infinity in the thermodynamic limit. However, in practical calculations, $L$ is typically finite. To account for this discreteness, we introduce a small parameter $\delta = 20t_c/L = 20 \epsilon_c/nN$ as a replacement for the infinitesimal $0^+$ in Eqs. (\ref{eq:S_ESE}), (\ref{eq:S_GSB}), and (\ref{eq:S_ESA}). This parameter helps eliminate the discretization effects in the single-particle energy levels. Thus, $\delta$ serves as an artificial resolution limit for resonance peaks in our calculations, indicating that only resonance widths larger than $\delta$ are considered physically meaningful. On the other hand, the resonance positions are generally not strongly affected by increasing $L$ beyond a certain threshold. It is important to note that our numerical simulations do not include any phenomenological parameters such as decoherence rates, which are often used for qualitative understanding of experimental data.

\begin{figure*}
\begin{centering}
\includegraphics[width=1.0\textwidth]{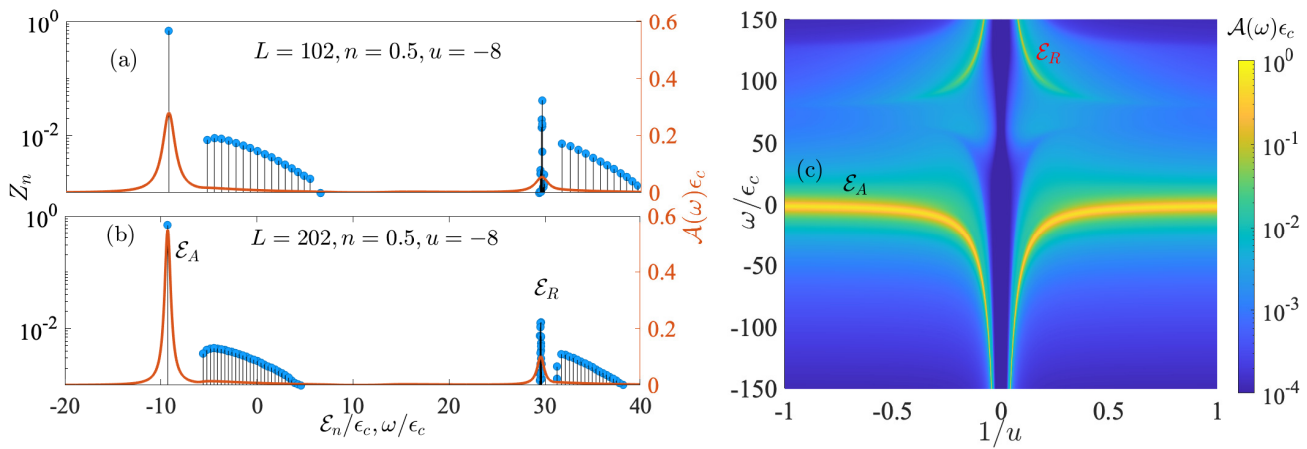}
\caption{\label{fig:1DAw} The spectral function of 1D polaron with parameters $n=0.5$, $t_c=4\epsilon_c$, and $t_d=4\epsilon_c$. (a) and (b) display the spectral function for two different lattice sizes, $L=102$ and $L=202$, respectively. The chosen interaction strength is $u=-8$. In addition, we include plots of the residue $Z_n$ as a function of the polaron energy $\mathcal E_n$ for comparison.(c) illustrates the spectral function as a function of the inverse interaction parameter $1/u$ with lattice size $L=150$. Please note that the color axis in (c) is displayed in a logarithmic scale.}
\end{centering}
\end{figure*}

Figure \ref{fig:1DAw} illustrates the spectral function for a single impurity, denoted as $\mathcal{A}(\omega)$, which represents the absorption spectra of our 1D polaron. It is defined as the negative imaginary part of the quantity $\sum_n{Z_n}/({\omega+i\delta-{\mathcal E}_n})$. In the figure, we have removed the trivial mean-field contribution by applying $\mathcal E_n\rightarrow \mathcal E_n-nU$. The parameters chosen for the plot are $n=0.5$, $t_c=4\epsilon_c$, and $t_d=4\epsilon_c$.

In Fig. \ref{fig:1DAw}(a) and (b), we present the spectral function at $u=-8$ for two different lattice sizes, $L=102$ and $L=202$, respectively. Two distinct polaron resonances can be observed, one at negative frequency $\mathcal E_A \approx -9.3 \epsilon_c$ and the other at positive frequency $\mathcal E_R \approx 29.6 \epsilon_c$. These resonances are commonly referred to as the attractive and repulsive polarons \cite{Massignan2014}, respectively. It can be noted that the resonance peaks have different widths for the two lattice sizes, while their positions are approximately the same. For comparison, we also plot the residue $Z_n$ as a function of $\mathcal E_n$. It can be seen that the attractive polaron resonance corresponds to a single polaron state with large residue, while the repulsive polaron receives contributions from multiple many-body eigenstates.

In Fig. \ref{fig:1DAw}(c), we examine the dependence of the spectral function on the interaction parameter $1/u$. It exhibits an interesting symmetry between positive and negative interactions, which arises due to the half-filling condition $n=N/L=0.5$ and the removal of mean-field contributions.

\begin{figure*}
\begin{centering}
\includegraphics[width=1.0\textwidth]{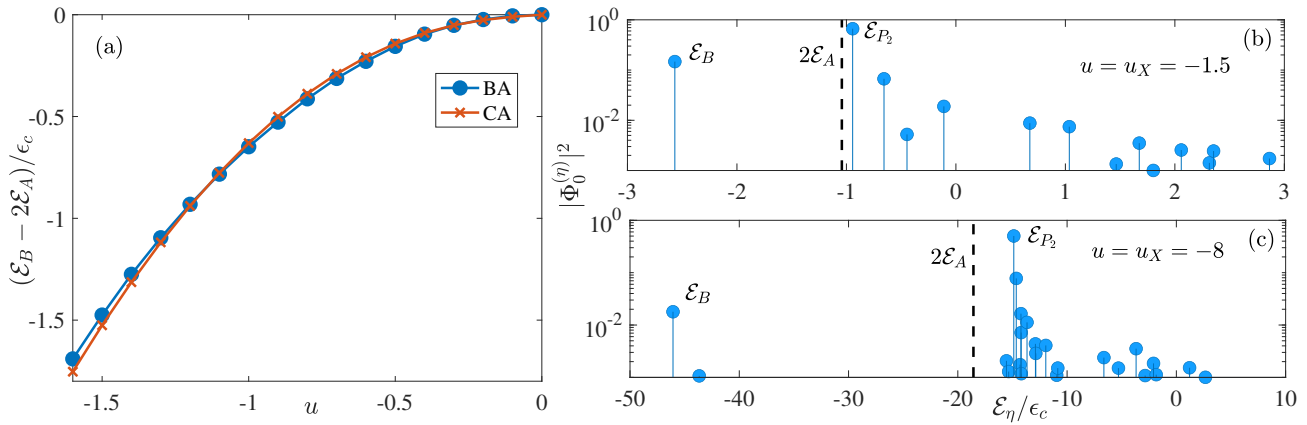}
\caption{\label{fig:EnBipol} (a) The comparison between our numerical results for the bipolaron binding energy using Chevy's ansatz (CA) with one particle-hole excitation and the exact results using Bethe's ansatz (BA) from Ref. \cite{Huber2019}. The CA calculation is carried out for $L=50$ and $N=25$. (b) and (c) shows the projection probability of bi-polaron to two free impurity $|\Phi_0^{(\eta)}|^2$ as a function of the corresponding many-body eigen energy $\mathcal E_\eta$ for parameter $u=u_X=-1.5$ and $u=u_X=-8$, respectively. The dashed line shows $2\mathcal E_A$ for comparison.}
\end{centering}
\end{figure*}

We have compared our numerical results for the two-polaron case using the ansatz in Eq. (\ref{eq:BipolaronAnsatz}) and parameters $L=50$, $N=25$ with exact Bethe's ansatz calculations from Ref. \cite{Huber2019} for the case of $u=u_X$. The results, shown in Figure \ref{fig:EnBipol} (a), demonstrate excellent agreement. The energy differences between the lowest $\mathcal{E}_\eta$ (denoted as $\mathcal{E}_B$) and two times the lowest $\mathcal{E}_n$ (corresponding to the attractive polaron energy $\mathcal{E}_A$) are displayed in this figure. The quantity $\mathcal{E}_B - 2\mathcal{E}_A$ represents the binding energy of two attractive polarons in the bipolaron ground state. Figures \ref{fig:EnBipol} (b) and (c) depict $|\Phi_0^{(\eta)}|^2$ as a function of $\mathcal{E}_\eta$ for $u=u_X=-1.5$ and $-8$, respectively. Here, $|\Phi_0^{(\eta)}|^2$ plays a similar role to $Z_n$ for single polaron states. It can be observed that $|\Phi_0^{(\eta)}|^2$ is small for the many-body ground state corresponding to $\mathcal{E}_B$, indicating that this state is bound and has little overlap with non-interacting scattering states. Conversely, another state with energy $\mathcal{E}_{P_2} > 2\mathcal{E}_A$ exhibits a large $|\Phi_0^{(\eta)}|^2$, as indicated in Figures \ref{fig:EnBipol} (b) and (c). We will see later that this $P_2$ state contributes significantly to $\mathcal{S}_{\rm EAE}$.

\subsection{Zero mixing time delay $t_{2}=0$}

\begin{figure*}
\begin{centering}
\includegraphics[width=1.0\textwidth]{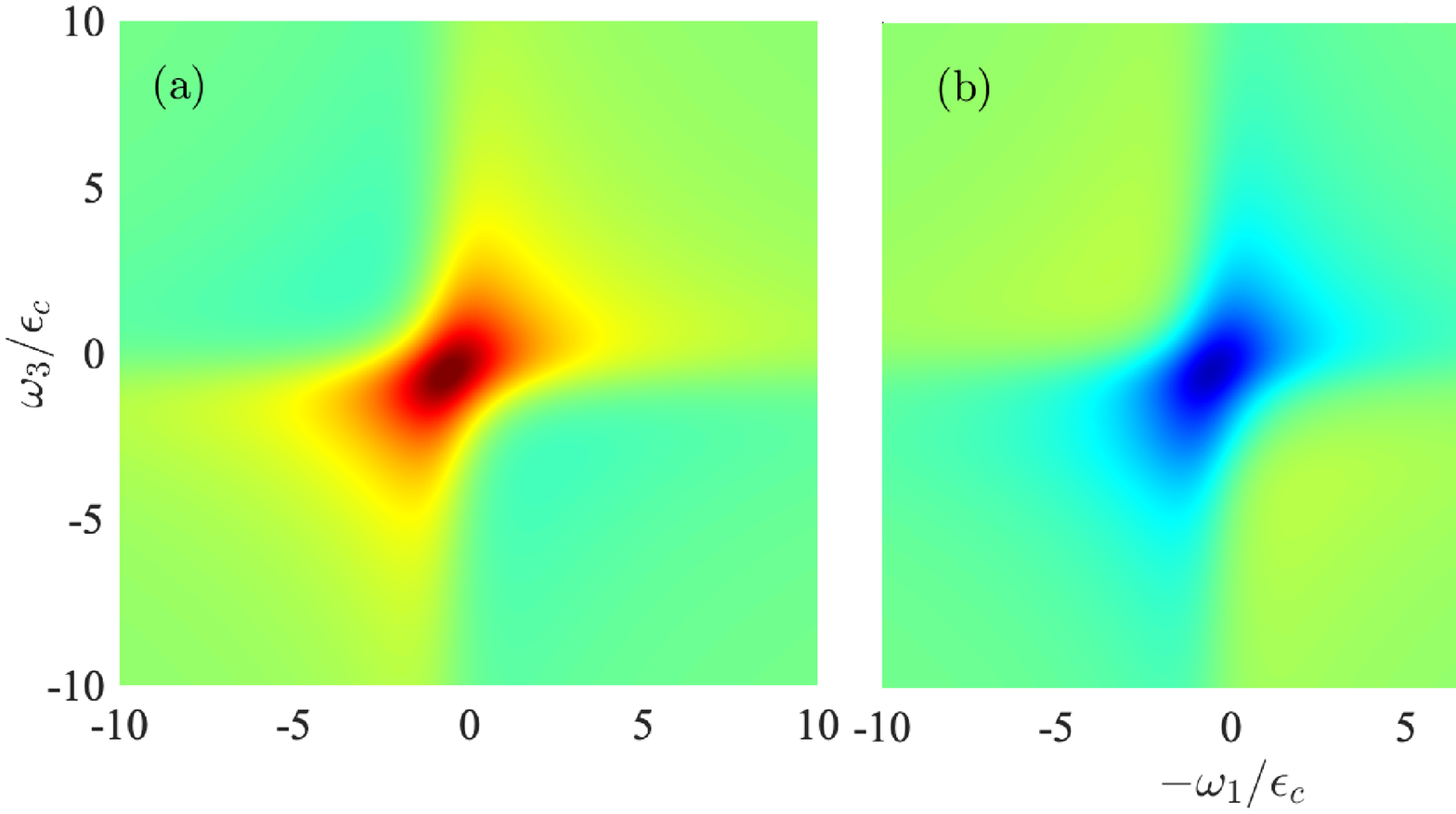}
\caption{\label{fig:S_u1p5} The 2DCS for $u=u_X=-1.5$, $L=50$, and $N=25$. (a) shows $\mathcal S_{\rm ESE}+\mathcal S_{\rm GSB}$ (b) shows $\mathcal S_{\rm ESA}$ and (c) shows the total.}
\end{centering}
\end{figure*}
\begin{figure*}
\begin{centering}
\includegraphics[width=1.0\textwidth]{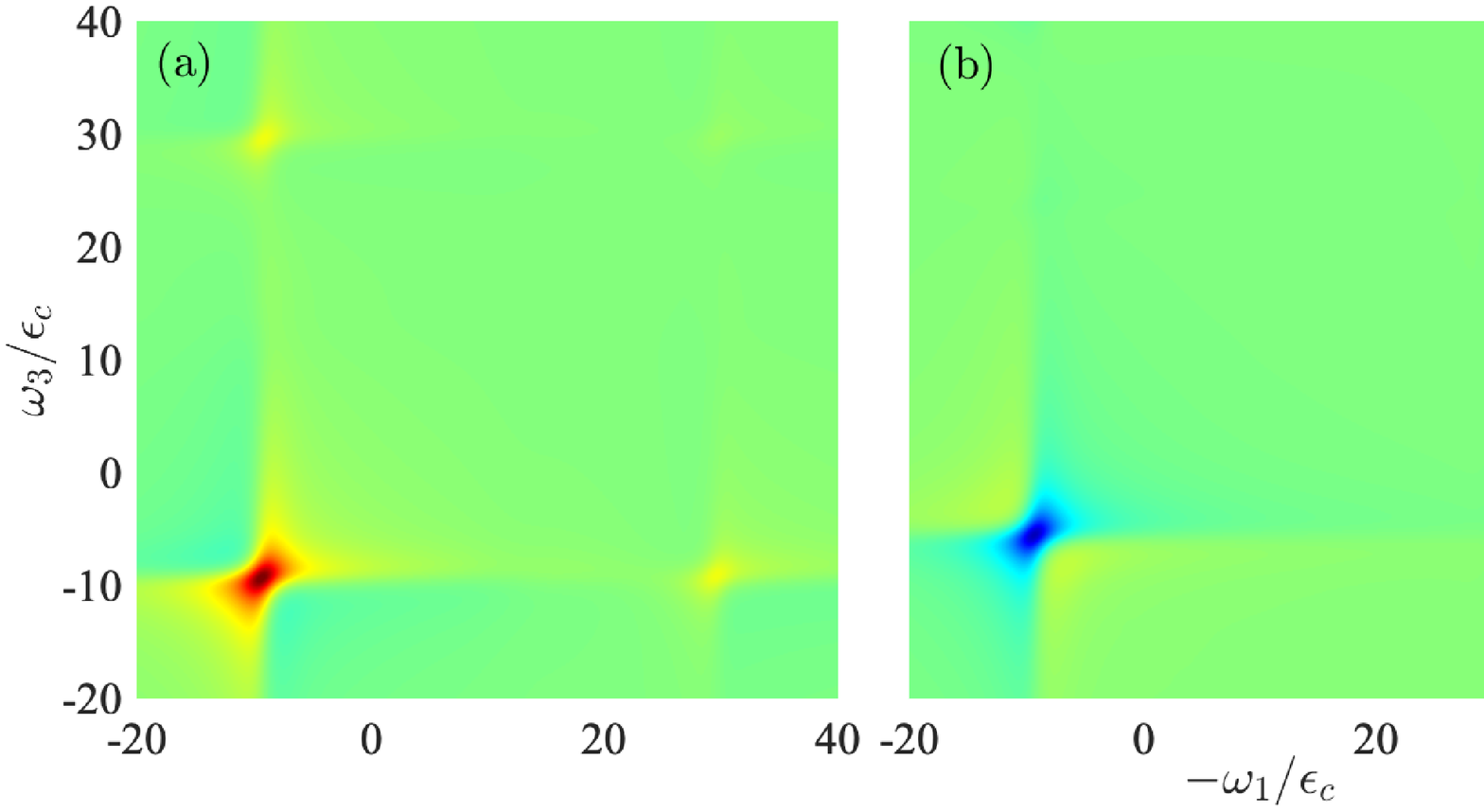}
\caption{\label{fig:S_u8} The 2DCS for $u=u_X=-8$, $L=50$, and $N=25$. (a) shows $\mathcal S_{\rm ESE}+\mathcal S_{\rm GSB}$ (b) shows $\mathcal S_{\rm ESA}$ and (c) shows the total.}
\end{centering}
\end{figure*}

Let us consider the case where the mixing time delay is zero, $t_2=0$. In this case, the expressions for the different contributions to the signal can be simplified as:
\begin{eqnarray}
\mathcal{S}_{\rm ESE}\left(\omega_{1},0,\omega_{3}\right)&=&\mathcal{S}_{\rm GSB}\left(\omega_{1},0,\omega_{3}\right) \nonumber \\
&=&-2\sum_{nm}\frac{Z_{n}}{\omega_{1}^{+}+\mathcal{E}_n}\frac{Z_{m}}{\omega_{3}^{+}-\mathcal{E}_m},\label{eq:S1_t2zero} \\
\mathcal{S}_{\rm EAS}\left(\omega_{1},0,\omega_{3}\right)&=&\sum_{n\eta}\frac{\phi_0^{(n)}}{\omega_1^++\mathcal E_n}\frac{\Phi^{(n,\eta)}\Phi_0^{(\eta)*}}{\omega_3^+-\mathcal E_\eta +\mathcal E_n}.\label{eq:S3_t2zero}
\end{eqnarray}
We present the numerical results for these quantities in Figs. \ref{fig:S_u1p5} and \ref{fig:S_u8} for the cases of interaction strength $u=u_X=-1.5$ and $u=u_X=-8$, respectively. In these figures, (a) shows ${\rm Re}[\mathcal{S}_{\rm ESE}+\mathcal{S}_{\rm GSB}]$, (b) shows ${\rm Re}[\mathcal{S}_{\rm ESA}]$, and (c) shows the real part of the total $\mathcal{S}_{\rm TOT}=\mathcal{S}_{\rm ESE}+\mathcal{S}_{\rm GSB}+\mathcal{S}_{\rm ESA}$.

For weak interaction $u=u_X=-1.5$, we observe that ${\rm Re}[\mathcal{S}_{\rm ESE}+\mathcal{S}_{\rm GSB}]$ in Fig. \ref{fig:S_u1p5} (a) is symmetric with respect to $-\omega_1$ and $\omega_3$, and is dominated by a peak around $(-\omega_1,\omega_3)\approx(\mathcal{E}_A,\mathcal{E}_A)$, as expected \cite{Hao2016,Hu2022arXiv}. On the other hand, ${\rm Re}[\mathcal{S}_{\rm ESA}]$ in Fig. \ref{fig:S_u1p5} (b) is dominated by a non-symmetric dip around $(-\omega_1,\omega_3)\approx(\mathcal{E}_A,\mathcal{E}_{P_2}-\mathcal{E}_A)$. From Fig. \ref{fig:EnBipol} (b), we can see that $\mathcal E_{P_2}\approx 2\mathcal{E}_{A}$, indicating $\mathcal{E}_{P_2}-\mathcal{E}_A\approx \mathcal{E}_A$. Therefore, the peak in Fig. \ref{fig:S_u1p5} (a) and the dip in (b) mostly cancel each other, resulting in a much weaker feature in the total signal shown in (c). This agrees with our expectation that weak interaction leads to weak nonlinear response.

For strong interaction $u=u_X=-8$, Fig. \ref{fig:S_u8} (a) shows diagonal peaks at $(-\omega_1,\omega_3)\approx(\mathcal{E}_A,\mathcal{E}_A)$ and $(\mathcal{E}_R,\mathcal{E}_R)$, corresponding to the attractive and repulsive polaron peaks. The off-diagonal peaks $(-\omega_1,\omega_3)\approx(\mathcal{E}_A,\mathcal{E}_R)$ and $(\mathcal{E}_A,\mathcal{E}_R)$ represent the quantum coherences between the polarons \cite{Hao2016,Tempelaar2019,Hu2022arXiv}. In Fig. \ref{fig:S_u8} (b), we observe that $\mathcal{S}_{\rm ESA}$ is dominated by a dip around $(-\omega_1,\omega_3)\approx(\mathcal{E}_A,\mathcal{E}_{P_2}-\mathcal{E}_A)$. However, the center of the dip deviates sufficiently from the peak at $(\mathcal{E}_A,\mathcal{E}_A)$, allowing both the peak and the dip structure to be observed in the total signal in (c).

It is interesting to note that the ground state of the two-impurity system at $\mathcal{E}_B$ does not contribute significantly, suggesting that it corresponds to a bound state with limited overlap with two free polarons. In contrast, the state associated with $\mathcal{E}_{P_2} \approx -14.8 \epsilon_c$ exhibits a large overlap, allowing us to interpret the difference $\mathcal{E}_{P_2} - 2\mathcal{E}_A$, which represents the distances between the dip and the peak in Fig. \ref{fig:S_u8} (c), as the induced interaction $U_{\rm int}$ between two free attractive polarons. Notably, this induced interaction $U_{\rm int}\approx 3.8 \epsilon_c$ is positive, and this behavior can be understood in terms of phase-space filling effects \cite{Muir2022}. Furthermore, it is worth mentioning that for our chosen parameter regime, $U_{\rm int}$ is much smaller than $\mathcal{E}_A$, which suggests that the interpretation of $U_{\rm int}$ as an interaction in a perturbative picture remains self-consistent. This interpretation can potentially be extended to interactions among multiple polarons.

\subsection{Mixing time $t_2$ dynamics}

\begin{figure*}
\begin{centering}
\includegraphics[width=1.0\textwidth]{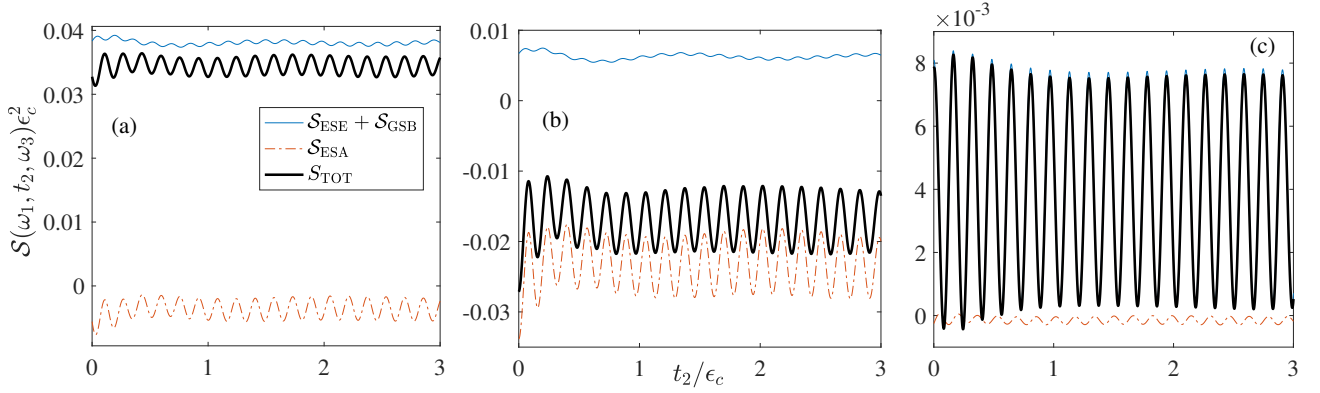}
\caption{\label{fig:t2dep} For the parameter $u=u_X=-8$, $L=50$, and $N=25$, this figure depicts the simulated real part of the rephasing 2D signal $\mathcal S(\omega_1,t_2,\omega_3)$ as a function of the mixing time delays $t_2$ at three specific regions: (a) the diagonal peak $(-\omega_1,\omega_3)\approx (\mathcal E_A, \mathcal E_A)$, (b) the dip at $(\mathcal E_A, \mathcal E_{P_2} - \mathcal E_A)$, and (c) the off-diagonal crosspeak $(\mathcal E_A, \mathcal E_{R})$. The thin blue solid and red dashed-dotted curves represent the contributions of $\mathcal S_{\rm ESE}+\mathcal S_{\rm GSB}$ and $\mathcal S_{\rm ESA}$, respectively. The thick black solid curve represents the total signal.}
\end{centering}
\end{figure*}

Figure \ref{fig:t2dep} illustrates the mixing time dynamics of the 2DCS at several different frequencies $(-\omega_1,\omega_3)$ for $u=u_X=-8$, $L=50$, and $N=25$. Panel (a) displays the diagonal peak corresponding to the attractive polaron around $(\mathcal E_A, \mathcal E_A)$, while panel (c) depicts the off-diagonal peak around $(\mathcal E_A, \mathcal E_R)$. The contribution to the total signal $\mathcal S_{\rm TOT}$ (thick black solid curve) for these two peaks primarily stems from $\mathcal S_{\rm ESE}+\mathcal S_{\rm GSB}$ (thin blue solid curve) throughout all times. In contrast, the total signal dip at $(\mathcal E_A, \mathcal E_{P_2} - \mathcal E_A)$ shown in panel (b) is predominantly determined by $\mathcal S_{\rm ESA}$ (thin red dash-dotted curve). We observe that all signals exhibit fast oscillations with a frequency of $\omega_2 = \mathcal E_R -\mathcal E_A \approx 38.9 \epsilon_c$. However, the behavior over longer time scales is less regular, possibly due to the finite lattice spacing.

\section{Conclusions}

In conclusions, we have presented a microscopic theory of the rephasing 2DCS of 1D exciton-polarons in monolayer TMD materials. Our theory includes the crucial excited-state absorption process, which is less considered in earlier theoretical investigations \cite{Tempelaar2019,Lindoy2022,Hu2022arXiv,Hu2023AB}. In the weak-coupling limit, this process cancels out the other two contributions from the excited-state emission process and the ground-state bleaching process. In the strong-coupling limit, it provides useful features to visualize the polaron-polaron interaction. We have carried out numerical calculations within Chevy's ansatz that includes one-particle-hole excitations of the Fermi sea. Our results are quantitatively reliable, as benchmarked by the exact Bethe ansatz solution for the bipolaron binding energy at the specific interaction parameter. Further improvement with multi-particle-hole excitations would provide independent confirmation of the accuracy of our predictions. Alternatively, future 2DCS measurements of exciton-polarons in 1D strain-engineered monolayer MoSe$_{2}$ and WSe$_{2}$ may present a stringent test of our results.

Our full microscopic calculations of the rephasing 2DCS of exciton-polarons can be easily extended to the two-dimensional configuration. However, the numerical effort becomes enormous due to the much-enlarged Hilbert space. A possible solution is to find out the most important intermediate states for bipolarons. This issue will be addressed in future studies. 

\section{Acknowledgements}
We would like to express our gratitude to Jesper Levinsen and Meera Parish for their inspiring suggestion to explore 1D systems. Additionally, we would like to thank Dmitry Efimkin and Jeffrey Davis for their stimulating discussion. This research was supported by the Australian Research Council's (ARC) Discovery Program, Grants No. DE180100592 and No. DP190100815 (J.W.), and Grant No. DP180102018 (X.-J.L).

\appendix
\section{Matrix elements of the model Hamiltonian}
In the extended Chevy's ansatz, Eq. (\ref{eq:BipolaronAnsatz}), which corresponds to the case of two bosonic impurities, the prime in the summation over $k$ indicates the avoidance of double counting. To be explicit, we define
\begin{equation}
\left|k\right\rangle_2 = C_kX^\dagger_{-k}X^\dagger_k\left|{\rm FS}\right \rangle,
\end{equation}
with the condition $k \ge 0$, and
\begin{equation}
\left| k k_p k_h\right\rangle_2 = C_{k,k_p,k_h}X_{-k}^{\dagger}X_{k-k_{p}+k_{h}}^{\dagger}c_{k_{p}}^{\dagger}c_{k_{h}}\left|\textrm{FS}\right\rangle
\end{equation}
with the restriction that $k-k_{p}+k_{h} \ge -k$. Here, $C_k=1/\sqrt{1+\delta_{k,0}}$ and $C_{k,k_p,k_h}=1/\sqrt{1+\delta_{-k,k_{\rm ph}}}$, where $\delta_{ab}$ is the Kronecker delta function, are introduced to ensure the orthogonality of the basis. We also define $k_{\rm ph} = k-k_p+k_h$ for convenience. It should be emphasized, as mentioned in the main text, that any momentum index appearing should be projected into the first Brillouin zone to impose the periodic boundary condition.

\begin{widetext}
The matrix elements of the Hamiltonian $\mathcal H_2$ expanded by $\left|k\right\rangle_2$ and $\left| k k_p k_h\right\rangle_2$ can be explicitly given by,
\begin{equation}
_2\left\langle k^{\prime}|H| k\right\rangle_2=\left[\left(E_{\mathrm{FS}}+\epsilon_k^X+\epsilon_{-k}^X+2 n U\right) \delta_{k^{\prime} k}+\frac{2 V}{L} C_{k^{\prime}} C_k\right],
\end{equation}
\begin{equation}
_2\left\langle k^{\prime}|H| k k_p k_h\right\rangle_2=\frac{U}{L}\left(\delta_{k^{\prime} k_{\rm ph}}+\delta_{k^{\prime},-k_{\rm ph}}+\delta_{k^{\prime},-k}+\delta_{k^{\prime} k}\right) C_{k^{\prime}} C_{k k_p k_n},
\end{equation}
and
\begin{eqnarray}
_2\left\langle k^{\prime}k_{p}^{\prime}k_{h}^{\prime}|H|kk_{p}k_{h}\right\rangle_2 & = & \left(E_{\mathrm{FS}}-\epsilon_{k_{h}}+\epsilon_{k_{p}}+\epsilon_{-k}^{X}+\epsilon_{k_{{\rm ph}}}^{X}+2nU\right)\delta_{k^{\prime}k}\delta_{k_{p}^{\prime}k_{p}}\delta_{k_{h}^{\prime}k_{h}} \nonumber \\
 & + &C_{k^{\prime}k_{p}^{\prime}k_{h}^{\prime}}C_{kk_{p}k_{h}}\frac{U}{L}(\delta_{k_{h}^{\prime}k_{h}}-\delta_{k_{p}^{\prime}k_{p}})\left(\delta_{k_{{\rm ph}}^{\prime}k_{{\rm ph}}}+\delta_{-k^{\prime},-k}+\delta_{-k^{\prime},k_{{\rm ph}}}+\delta_{k_{{\rm ph}}^{\prime},-k}\right) \nonumber \\
 & + &C_{k^{\prime}k_{p}^{\prime}k_{h}^{\prime}}C_{kk_{p}k_{h}}\frac{2V}{L}\delta_{k_{p}^{\prime}k_{p}}\delta_{k_{h}^{\prime}k_{h}},
\end{eqnarray}
where in practical calculations, we often subtract the unimportant Fermi sea energy $E_{\rm FS}$ and the mean-field contribution $2nU$ from the diagonal matrix elements.

For completeness, we also list here the matrix elements of $\mathcal H_1$ expanded by the basis $\left|0\right\rangle_1$ and $\left|k_pk_h\right\rangle_1$:
\begin{equation}
_1\langle0|H|0\rangle_1=E_{{\rm FS}}+nU+\epsilon_{k=0}^{X}, 
\end{equation}
\begin{equation}
_1\langle0|H|k_{p}k_{h}\rangle_1=\frac{U}{L}, 
\end{equation}
\begin{equation}
_1\langle k_{p}k_{h}|H|k_{p}^{\prime}k_{h}^{\prime}\rangle_1=[E_{{\rm FS}}+nU+\Delta\epsilon(k_{p},k_{h})]\delta_{k_{p}k_{p}^{\prime}}\delta_{k_{h}k_{h}^{\prime}}+\frac{U}{L}(\delta_{k_{h}k_{h}^{\prime}}-\delta_{k_{p}k_{p}^{\prime}}),
\end{equation}
where $\Delta\epsilon(k_{p},k_{h})=\epsilon_{k_{p}}-\epsilon_{k_{h}}+\epsilon_{-k_{p}+k_{h}}^{X}$.
\end{widetext}

\begin{figure}
\begin{centering}
\includegraphics[width=0.98\columnwidth]{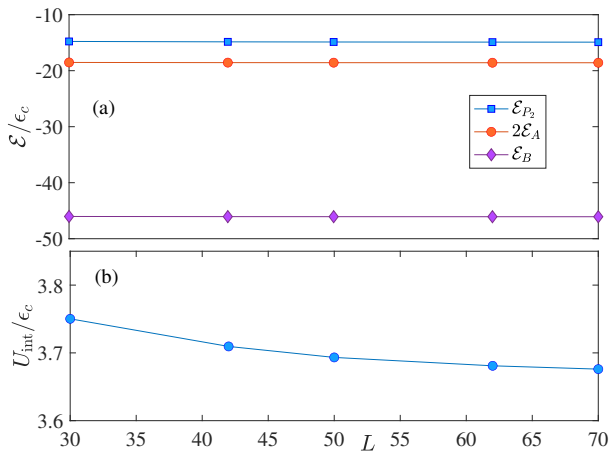}
\caption{\label{fig:EvsL} (a) $\mathcal E_{P_2}$, $2 \mathcal E_A$ and $\mathcal E_B$ as a function of $L$, all of which converge to constant in large $L$ limit. (b) $U_{\rm int} = \mathcal E_{P_2}-2\mathcal E_A$ as a function of $L$, which converges to a positive constant value in large $L$ limit. Other parameters are $n=0.5$ and $u=u_X=-8$.}
\end{centering}
\end{figure}

\section{Finite-size scaling for the bipolaron states}
It is important to investigate the thermodynamic limit $L\rightarrow \infty$, $N\rightarrow \infty$ and $n = N/L\rightarrow 0.5$. Figure \ref{fig:EvsL} shows that the quantities $\mathcal E_{P_2}$, $\mathcal E_{A}$, $\mathcal E_{B}$, and $U_{\rm int} = \mathcal E_{P_2}-2 \mathcal E_A$ converge to constant in this limit.

\bibliography{RefMDCS_FPolaron1D}

\end{document}